\documentclass[twocolumn,showpacs,preprintnumbers,amsmath,amssymb]{revtex4}
\usepackage{graphicx}% Include figure files
\usepackage{dcolumn}% Align table columns on decimal point
\usepackage{bm}% bold math

\begin{document}

\preprint{ }

\title{{\bf  Reply to Comment on "Quantum dense key distribution"  }}

\author{I. P. Degiovanni}
 \email{degio@ien.it}
\author{I. Ruo Berchera}
\author{S. Castelletto}
\author{M. L. Rastello }
\affiliation{Istituto Elettrotecnico Nazionale G. Ferraris \\
Strada delle Cacce 91-10135 Torino (Italy)}

\author{F. A. Bovino }%
\author{A. M. Colla}
\author{G. Castagnoli}

\affiliation{%
ELSAG SpA \\
Via Puccini 2-16154 Genova (Italy)
}%

\date{\today}

\begin{abstract}

In this Reply we propose a modified security proof of the Quantum
Dense Key Distribution protocol detecting also the eavesdropping
attack proposed by W\'{o}jcik in his Comment.

\end{abstract}


\maketitle

%\draft

%\setlength{\baselineskip}{5ex}

In his Comment \cite{wojcikcomment}, W\'{o}jcik  proposes two
simple schemes for performing eavesdropping attacks to the Quantum
Dense Key Distribution protocol (QDKD) in Ref. \cite{qdkd}. These
schemes enable Eve to achieve a larger mutual information than
Alice's and Bob's one, but maintaining inviolated the security
condition (Eq. (6) in Ref. \cite{qdkd}).

Eve's attack in W\'{o}jcik's schemes relies on the possibility of
subtraction of photons from the quantum channel without being
disclosed by the Anticorrelation Check. In fact the security proof
given in \cite{qdkd} is incomplete because it includes only the
effect that Eve's presence induces a bit flip on the travelling
qubit.

In this reply we complete the security proof of QDKD even against
individual eavesdropping attack with injection or subtraction of
photons in the quantum channels. Furthermore we discuss the
security of the experimental realization of the QDKD protocol
performed in \cite{qdkd}, according to the arguments raised by the
modified proof.

A dedicated formalism is introduced to account for these attacks.
Let $|m^{n} _{\mathrm{X}}\rangle $ be the state with $n$ photons
on the quantum channel X, where $m$ photons have horizontal
polarization and $(n-m)$ vertical polarization. The orthonormal
base $\mathcal{B}_{\mathrm{X}}:\{|m^{n} _{\mathrm{X}}\rangle \}$
spans the Hilbert space $\mathcal{H}_{\mathrm{X}}$ of photons in
channel X; $|0^{0} _{\mathrm{X}}\rangle $ is the vacuum state.

In the actual formalism, referring to Fig. 1 in Ref. \cite{qdkd},
Alice produces pairs of photons in the singlet state
$|\psi^{-}_{\mathrm{AB}}\rangle=\frac{1}{\sqrt{2}}(|0^{1}
_{\mathrm{A}}1^{1} _{\mathrm{B}}\rangle -|1^{1}_{\mathrm{A}} 0^{1}
 _{\mathrm{B}}\rangle)$. Photon A is stored in her laboratory while on photon B
she performs either the operation $\mathbf{1}_{\mathrm{B}}$
(identity operator) or $\widehat{Z}_{\mathrm{B}}$ before sending
it to Bob. A generalized gate $\widehat{Z}_{\mathrm{B}}$ acts in
the Hilbert space $\mathcal{H}_{\mathrm{B}}$ as
$\widehat{Z}_{\mathrm{B}}|m^{n} _{\mathrm{B}}\rangle
=(-1)^{m}|m^{n} _{\mathrm{B}}\rangle $. $\widehat{Z}_{\mathrm{B}}$
is hermitian and unitary, and corresponds to the Pauli matrix
$\widehat{\sigma}_{Z}$ acting individually on each photon in the
channel B. Alice's selection of gate $\mathbf{1}_{\mathrm{B}}$ or
$\widehat{Z}_{\mathrm{B}}$ is encoded according to
\begin{eqnarray}
\mathbf{1}_{\mathrm{B}}|\psi^{-}_{\mathrm{AB}}\rangle&=&|\psi^{-}_{\mathrm{AB}}\rangle
\; \longrightarrow \; \mathrm{bit } \; 0  \nonumber \\
\widehat{Z}_{\mathrm{B}}|\psi^{-}_{\mathrm{AB}}\rangle&=&-|\psi^{+}_{\mathrm{AB}}
\rangle \; \longrightarrow \; \mathrm{bit } \; 1 , \label{a1}
\end{eqnarray}
with $|\psi^{+}_{\mathrm{AB}}\rangle=\frac{1}{\sqrt{2}}(|0^{1}
_{\mathrm{A}}1^{1}_{\mathrm{B}} \rangle +|1^{1}_{\mathrm{A}} 0^{1}
_{\mathrm{B}} \rangle )$.

Bob randomly switches photon B towards either the Anticorrelation
Check or his encoding apparatus.

The Anticorrelation Check is performed by Bob projecting photon B
on the states $|0^{1}_{\mathrm{B}}\rangle$ and
$|1^{1}_{\mathrm{B}}\rangle$ and Alice projecting photon A on the
states $|1^{1}_{\mathrm{A}}\rangle,$ and $
|0^{1}_{\mathrm{A}}\rangle$, respectively. The non-local
measurement guarantees the security of the transmission.

Bob's encoding apparatus is identical to Alice's. The
communication takes place sending back photon B to Alice, who
performs the incomplete Bell's state analysis. Specifically,
$|\psi^{+}_{\mathrm{AB}}\rangle$ corresponds to Alice and Bob
encoding 0 and 1 or 1 and 0 respectively, while
$|\psi^{-}_{\mathrm{AB}}\rangle$ corresponds to  Alice and Bob
both encoding 0 or 1. In other words the measurement of
$|\psi^{+}_{\mathrm{AB}}\rangle$, $|\psi^{-}_{\mathrm{AB}}\rangle$
corresponds to the sum mod 2 of the bits encoded by Alice and Bob.

We model the general individual Eve's attack by coupling photon B
with an \textit{ancilla} system of Hilbert space
$\mathcal{H}_{\mathrm{E}}$ in the initial state
$|e_{\mathrm{E}}\rangle$ by means of general unitary operators
$\widehat{J}_{\mathrm{BE}}$ and $\widehat{K}_{\mathrm{BE}}$ before
and after Bob's operations, respectively. The final state belongs
to the widened Hilbert space $\mathcal{H}_{\mathrm{A}} \otimes
\mathcal{H}_{\mathrm{B}} \otimes \mathcal{H}_{\mathrm{E}}$. This
approach is general because any physical non-unitary interaction
is equivalent to a unitary one with a higher dimensional ancilla
space \cite{gisinperes}.

The final state after Bob's and Eve's operations is described by
the trace preserving quantum operation $\mathcal{E}$
\begin{equation}\label{ee}
\mathcal{E}(\widehat{\rho}_{\mathrm{AB}})
=\frac{1}{2}\mathcal{E}_{\mathbf{1}_{\mathrm{B}}}(\widehat{\rho}_{\mathrm{AB}})
+\frac{1}{2}
\mathcal{E}_{\widehat{Z}_{\mathrm{B}}}(\widehat{\rho}_{\mathrm{AB}}),
\end{equation}
where $\mathcal{E}_{\widehat{Z}_{\mathrm{B}}}$ and
$\mathcal{E}_{\mathbf{1}_{\mathrm{B}}}$ are quantum operations
describing the evolution of the initial state
$\widehat{\rho}_{\mathrm{AB}}$ prepared by Alice and modified by
Bob's and Eve's actions
\begin{eqnarray}
\mathcal{E}_{\mathbf{1}_{\mathrm{B}}}(\widehat{\rho}_{\mathrm{AB}})&=&
\widehat{K}_{\mathrm{BE}}\mathbf{1}_{\mathrm{B}}\widehat{J}_{\mathrm{BE}}
\widehat{\rho}_{\mathrm{AB}}\otimes |e_{\mathrm{E}}\rangle\langle
e_{\mathrm{E}}|
\widehat{J}_{\mathrm{BE}}^{\dag}\mathbf{1}_{\mathrm{B}}\widehat{K}^{\dag}_{\mathrm{BE}},
\nonumber \\
\mathcal{E}_{\widehat{Z}_{\mathrm{B}}}(\widehat{\rho}_{\mathrm{AB}})&=&
\widehat{K}_{\mathrm{BE}}\widehat{Z}_{\mathrm{B}}\widehat{J}_{\mathrm{BE}}
\widehat{\rho}_{\mathrm{AB}}\otimes |e_{\mathrm{E}}\rangle\langle
e_{\mathrm{E}}|
\widehat{J}_{\mathrm{BE}}^{\dag}\widehat{Z}_{\mathrm{B}}\widehat{K}^{\dag}_{\mathrm{BE}}.
\label{e1ez}
\end{eqnarray}
It is assumed that Bob encodes bit 0 or 1 with probability 1/2.

Our aim is to quantify the maximum information achievable by Eve
in terms of the quantities measured by Alice and Bob in the
Anticorrelation Check. We define the quantities $\mathrm{P}_{01}$
and $\mathrm{P}_{10}$ as the probabilities of anticorrelated
results according to
\begin{eqnarray}
  \mathrm{P}_{01}=\mathrm{tr}[\widehat{J}_{\mathrm{BE}}\widehat{\rho}_{\mathrm{AB}}\otimes
  |e_{\mathrm{E}}\rangle\langle e_{\mathrm{E}}|\widehat{J}^{\dag}_{\mathrm{BE}}
  \widehat{\Pi}_{01}],   \nonumber \\
  \mathrm{P}_{10}=\mathrm{tr}[\widehat{J}_{\mathrm{BE}}\widehat{\rho}_{\mathrm{AB}}\otimes
  |e_{\mathrm{E}}\rangle\langle e_{\mathrm{E}}|\widehat{J}^{\dag}_{\mathrm{BE}}
  \widehat{\Pi}_{10}],  \label{panticorrs}
\end{eqnarray}
where $\widehat{\Pi}_{01}=|0^{1}_{\mathrm{A}}
1^{1}_{\mathrm{B}}\rangle  \langle
0^{1}_{\mathrm{A}}1^{1}_{\mathrm{B}}| $,
$\widehat{\Pi}_{10}=|1^{1}_{\mathrm{A}}0^{1}_{\mathrm{B}}\rangle
\langle 1^{1}_{\mathrm{A}}0^{1}_{\mathrm{B}}| $ are projection
operators. We assume $\widehat{\rho}_{\mathrm{AB}}=1/2
|\psi^{-}_{\mathrm{AB}}\rangle
  \langle \psi^{-}_{\mathrm{AB}} | + 1/2
|\psi^{+}_{\mathrm{AB}}\rangle  \langle \psi^{+} _{\mathrm{AB}}|$
(Alice prepares only states $|\psi^{\pm}_{\mathrm{AB}}\rangle$
with probability $\frac{1}{2}$), in the absence of Eve's attack
$\mathrm{P}_{01}=\mathrm{P}_{10}=0.5$ (perfect anticorrelation).
Eve's actions lower the value of $\mathrm{P}_{01}$ and
$\mathrm{P}_{10}$ and this is basically the signature of her
presence.

The maximum of the mutual information between Bob and Eve
$I_{\mathrm{B:E}}$, i.e. Eve's ability to distinguish Bob's
operations, can be evaluated exploiting the Holevo bound
\cite{NC00}: $I_{\mathrm{B:E}}\leq \mathcal{I}_{\mathrm{B:E}}$.
Consider the following scenario: Alice prepares the state
$\widehat{\rho}_{\mathrm{AB}}$, Bob encodes his key, and Eve
couples her system to photon B. The maximum mutual information
between Bob and Eve $I_{\mathrm{B:E}}$ is bounded by
\begin{equation}\label{ibee}
\mathcal{I}_{\mathrm{B:E}} =
S[\mathcal{E}(\widehat{\rho}_{\mathrm{AB}})] -\frac{1}{2}
S[\mathcal{E}_{\mathbf{1}_{\mathrm{B}}}(\widehat{\rho}_{\mathrm{AB}})]
-\frac{1}{2}
S[\mathcal{E}_{\widehat{Z}_{\mathrm{B}}}(\widehat{\rho}_{\mathrm{AB}})],
\end{equation}
where $S(\widehat{\rho})$ is the Von Neumann entropy \cite{NC00}
of the generic state $\widehat{\rho}$.

Analogously, also the maximum of mutual information between Alice
and Eve $I_{\mathrm{A:E}}$, i.e. Eve's ability to distinguish the
states prepared by Alice, is calculated exploiting the Holevo
bound ($I_{\mathrm{A:E}}\leq \mathcal{I}_{\mathrm{A:E}}$). In this
case
\begin{equation}\label{iaee}
  \mathcal{I}_{\mathrm{A:E}}=
  S[\mathcal{E}(\widehat{\rho}_{\mathrm{AB}})]
  -\frac{1}{2}S[\mathcal{E}(|\psi^{-}_{\mathrm{AB}} \rangle \langle \psi^{-} _{\mathrm{AB}}|)]
  -\frac{1}{2}S[\mathcal{E}(|\psi^{+}_{\mathrm{AB}}\rangle \langle \psi^{+}
  _{\mathrm{AB}}|)].
\end{equation}

To evaluate $\mathcal{I}_{\mathrm{A:E}}$ and
$\mathcal{I}_{\mathrm{B:E}}$ in terms of $\mathrm{P}_{01}$ and
$\mathrm{P}_{10}$ we define the final states in the coupled space
$\mathcal{H}_{\mathrm{A}} \otimes \mathcal{H}_{\mathrm{B}} \otimes
\mathcal{H}_{\mathrm{E}}$ after Bob's and Eve's operations
\begin{eqnarray}
\widehat{K}_{\mathrm{BE}}\mathbf{1}_{\mathrm{B}}\widehat{J}_{\mathrm{BE}}
|\psi^{+}_{\mathrm{AB}}\rangle\otimes|e_{\mathrm{E}}\rangle&=&|\mu^{+}_{\mathrm{ABE}}\rangle,
  \nonumber \\
\widehat{K}_{\mathrm{BE}}\mathbf{1}_{\mathrm{B}}\widehat{J}_{\mathrm{BE}}
|\psi^{-}_{\mathrm{AB}}\rangle\otimes|e_{\mathrm{E}}\rangle&=&|\mu^{-}_{\mathrm{ABE}}\rangle,
  \nonumber \\
\widehat{K}_{\mathrm{BE}}\widehat{Z}_{\mathrm{B}}\widehat{J}_{\mathrm{BE}}
|\psi^{+}_{\mathrm{AB}}\rangle\otimes|e_{\mathrm{E}}\rangle&=&|\nu^{+}_{\mathrm{ABE}}\rangle,
  \nonumber \\
\widehat{K}_{\mathrm{BE}}\widehat{Z}_{\mathrm{B}}\widehat{J}_{\mathrm{BE}}
|\psi^{-}_{\mathrm{AB}}\rangle\otimes|e_{\mathrm{E}}\rangle&=&|\nu^{-}_{\mathrm{ABE}}\rangle,
\label{jkact}
\end{eqnarray}
where we observe that $\langle \mu^{-}_{\mathrm{ABE}}|
\mu^{+}_{\mathrm{ABE}}\rangle=0$, $\langle \nu^{-}_{\mathrm{ABE}}|
\nu^{+}_{\mathrm{ABE}}\rangle=0$.

Before the Anticorrelation Check (or the Bob's operation) the
evolution of the system can be completely described by Eve's
operation of coupling photon B with her \textit{ancilla} system
\begin{eqnarray}
\widehat{J}_{\mathrm{BE}}|1^{1}_{\mathrm{B}}\rangle\otimes|e_{\mathrm{E}}\rangle&=&
\alpha|1^{1}_{\mathrm{B}}\rangle\otimes|\alpha_{\mathrm{E}}\rangle+
\gamma|\Gamma_{\mathrm{BE}}\rangle,
  \nonumber \\
\widehat{J}_{\mathrm{BE}}|0^{1}_{\mathrm{B}}\rangle\otimes|e_{\mathrm{E}}\rangle&=&
\beta|0^{1}_{\mathrm{B}}\rangle\otimes|\beta_{\mathrm{E}}\rangle+
\delta|\Delta_\mathrm{BE}\rangle, \label{jact}
\end{eqnarray}
with $\langle 1^{1}_{\mathrm{B}} |\Gamma_{\mathrm{BE}}\rangle=0$
and $\langle 0^{1}_{\mathrm{B}} |\Delta_{\mathrm{BE}}\rangle=0$.
$|\alpha_{\mathrm{E}}\rangle$, $|\beta_{\mathrm{E}}\rangle$,
$|\Gamma_{\mathrm{BE}}\rangle$, $|\Delta_{\mathrm{BE}}\rangle$ are
normalized to one and the operator $\widehat{J}_{\mathrm{BE}}$ is
unitary thus
$|\alpha|^{2}+|\gamma|^{2}=|\beta|^{2}+|\delta|^{2}=1$. The states
$|\Gamma_{\mathrm{BE}}\rangle$ and $|\Delta_{\mathrm{BE}}\rangle$
represent situations in which the Anticorrelation Check produces
unexpected ("wrong") results due to e.g. bit-flip, vacuum state or
state with more than one photon in the channel B. This is the main
difference with respect to the security proof proposed in
\cite{qdkd} where Alice and Bob considered that only the bit-flip
was the signature of Eve's presence. In this respect the
probabilities of anticorrelated results are
$\mathrm{P}_{01}=|\alpha|^{2}/2$ and
$\mathrm{P}_{10}=|\beta|^{2}/2$.

Thus, we can obtain the relation between $\mathrm{P}_{01}$ and
$\mathrm{P}_{10}$, and the final states of the coupled system, by
inserting Eq.s (\ref{jact}) in the left hand side of Eq.s
(\ref{jkact}). Observing that $\langle
\mu^{+}_{\mathrm{ABE}}|\nu^{-}_{\mathrm{ABE}}\rangle= \langle
\mu^{-}_{\mathrm{ABE}}|\nu^{+}_{\mathrm{ABE}}\rangle \triangleq
p$, $\langle
\mu^{+}_{\mathrm{ABE}}|\nu^{+}_{\mathrm{ABE}}\rangle=\langle
\mu^{-}_{\mathrm{ABE}}|\nu^{-}_{\mathrm{ABE}}\rangle\triangleq q$
we obtain
\begin{eqnarray}
p= \frac{c-d}{2} - \mathrm{P}_{01}(1+c)- \mathrm{P}_{10}(1-d),
  \nonumber \\
q= \frac{c+d}{2} - \mathrm{P}_{01}(1+c)+ \mathrm{P}_{10}(1-d),
 \label{relaz}
\end{eqnarray}
where $c=\langle \Gamma_{\mathrm{BE}}|\widehat{Z}_{\mathrm{B}}
|\Gamma_{\mathrm{BE}}\rangle$ and $d=\langle
\Delta_{\mathrm{BE}}|\widehat{Z}_{\mathrm{B}}
|\Delta_{\mathrm{BE}}\rangle$ are two real parameters ($ c$ and $d
\in [ -1, 1  ] $) under Eve's control that cannot be evaluated by
Alice and Bob \cite{note}.

\begin{figure}[tbp]
%[htbp]
\par
\begin{center}
\includegraphics[angle=0, width=9 cm, height=6 cm]{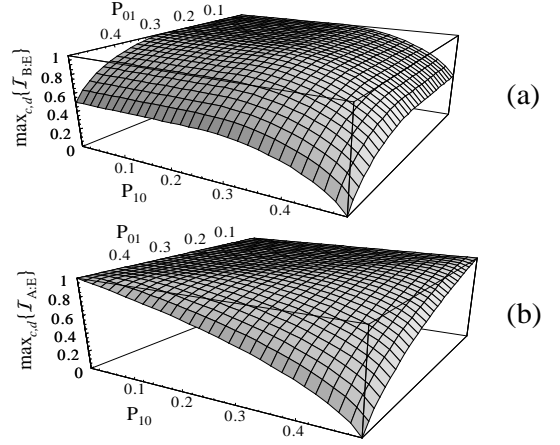}
\end{center}
\caption{ Plot of the maximum of $\mathcal{I}_{\mathrm{B:E}}$ (a)
and $\mathcal{I}_{\mathrm{A:E}}$ (b) versus $\mathrm{P}_{01}$ and
$\mathrm{P}_{10}$ } \label{Figure 2}
\end{figure}

According to Eq. (\ref{ibee}), we evaluate
$\mathcal{I}_{\mathrm{B:E}}$. From Eq.s (\ref{e1ez}) and Eq.s
(\ref{jkact}) we obtain
$S[\mathcal{E}_{\mathbf{1}_{\mathrm{B}}}(\widehat{\rho}_{\mathrm{AB}})]=1$,
$S[\mathcal{E}_{\widehat{Z}_{\mathrm{B}}}(\widehat{\rho}_{\mathrm{AB}})]=1$.
The calculation of $S[\mathcal{E}(\widehat{\rho}_{\mathrm{AB}})]$
is not trivial. In order to obtain the diagonal representation of
the state $\mathcal{E}(\widehat{\rho}_{\mathrm{AB}})$ we introduce
the orthonormal base $\mathcal{S}:\{|
\mu^{-}_{\mathrm{ABE}}\rangle, | \mu^{+}_{\mathrm{ABE}}\rangle, |
\xi^{(1)}_{\mathrm{ABE}}\rangle, | \xi^{(2)}_{\mathrm{ABE}}\rangle
\}$. $\mathcal{S}$ spans the generic subspace of the Hilbert space
$\mathcal{H}_{\mathrm{A}} \otimes \mathcal{H}_{\mathrm{B}} \otimes
\mathcal{H}_{\mathrm{E}}$ support of
$\mathcal{E}(\widehat{\rho}_{\mathrm{AB}})$. According to Eq.s
(\ref{relaz}), the states $| \nu^{-}_{\mathrm{ABE}}\rangle$ and $|
\nu^{+}_{\mathrm{ABE}}\rangle$ can be rewritten as
\begin{eqnarray}
| \nu^{-}_{\mathrm{ABE}}\rangle&=& p |
\mu^{+}_{\mathrm{ABE}}\rangle + q | \mu^{-}_{\mathrm{ABE}}\rangle
+ s | \xi^{(1)}_{\mathrm{ABE}}\rangle + t |
\xi^{(2)}_{\mathrm{ABE}}\rangle,  \nonumber \\
| \nu^{+}_{\mathrm{ABE}}\rangle&=& q |
\mu^{+}_{\mathrm{ABE}}\rangle + p | \mu^{-}_{\mathrm{ABE}}\rangle
+ r | \xi^{(1)}_{\mathrm{ABE}}\rangle, \nonumber
\end{eqnarray}
where $r$, $s$ and $t$ are complex, and $p$ and $q$, according
with Eq. (\ref{relaz}), are real. From the normalization and
orthogonality conditions on $| \nu^{+}_{\mathrm{ABE}}\rangle$ and
$| \nu^{-}_{\mathrm{ABE}}\rangle$ we obtain the Von Neumann
entropy of $\mathcal{E}(\widehat{\rho}_{\mathrm{AB}})$ as
\begin{equation}\label{ent1}
 S[\mathcal{E}(\widehat{\rho}_{\mathrm{AB}})]=-\sum_{i=1}^{4} \lambda_{i} \mathrm{log}
 \lambda_{i}.
\end{equation}
where $\lambda_{1}=\frac{1}{4}(1-p-q)$,
$\lambda_{2}=\frac{1}{4}(1+p+q)$, $\lambda_{3}=\frac{1}{4}(1-p+q)$
and $\lambda_{4}=\frac{1}{4}(1+p-q)$. Thus,
\begin{equation}\label{ibeeres}
 \mathcal{I}_{\mathrm{B:E}} = -\sum_{i=1}^{4} \lambda_{i} \mathrm{log}
 \lambda_{i}-1.
\end{equation}
As Alice and Bob have only access to the results of the
Anticorrelation Check, for any fixed pair of values
$\mathrm{P}_{01}$ and $\mathrm{P}_{10}$ the maximum information
achievable by Eve, $\max_{c,d} \{\mathcal{I}_{\mathrm{B:E}}\}$,
corresponds to the maximum value of $\mathcal{I}_{\mathrm{B:E}}$
in the range of values allowed for $c$ and $d$. As shown in Fig. 1
(a) the behavior of $\max_{c,d} \{\mathcal{I}_{\mathrm{B:E}}\}$
versus $\mathrm{P}_{01}$ and $\mathrm{P}_{10}$ can be analyzed by
considering four regions. Specifically
\begin{itemize}
  \item for $\mathrm{P}_{01}\geq 0.25$ and $\mathrm{P}_{10} \geq 0.25$
   \\
$\max_{c,d}
\{\mathcal{I}_{\mathrm{B:E}}\}=\mathcal{I}_{\mathrm{B:E}}(\mathrm{P}_{01},
\mathrm{P}_{10}, c=1, d=-1 )$
  \item for $\mathrm{P}_{01} < 0.25$ and $\mathrm{P}_{10} \geq 0.25$
  \\
$\max_{c,d}
\{\mathcal{I}_{\mathrm{B:E}}\}=\mathcal{I}_{\mathrm{B:E}}(\mathrm{P}_{01},
\mathrm{P}_{10}, c=\frac{2 \mathrm{P}_{01}}{1-2 \mathrm{P}_{01}},
d=-1 )$
  \item for $\mathrm{P}_{01}\geq 0.25$ and $\mathrm{P}_{10} < 0.25$
   \\
$\max_{c,d}
\{\mathcal{I}_{\mathrm{B:E}}\}=\mathcal{I}_{\mathrm{B:E}}(\mathrm{P}_{01},
\mathrm{P}_{10}, c=1, d=\frac{2 \mathrm{P}_{10}}{2
\mathrm{P}_{10}-1} ) )$
  \item for $\mathrm{P}_{01} < 0.25$ and $\mathrm{P}_{10} < 0.25$
 \\
$\max_{c,d} \{\mathcal{I}_{\mathrm{B:E}}\}=1$.
\end{itemize}
Thus, for $\mathrm{P}_{01}\geq 0.25$ or $\mathrm{P}_{10}\geq 0.25$
the maximum information achievable by Eve is upper-bounded, and
for $\mathrm{P}_{01}=\mathrm{P}_{10}=0.5$ (perfect
anticorrelation) Eve can get no information at all.

According to Eq. (\ref{iaee}), we evaluate
$\mathcal{I}_{\mathrm{A:E}}$. Observing that the eigenvalues of
both $\mathcal{E}(|\psi^{-}_{\mathrm{AB}} \rangle \langle \psi^{-}
_{\mathrm{AB}}|)$ and $\mathcal{E}(|\psi^{+}_{\mathrm{AB}}\rangle
\langle \psi^{+}_{\mathrm{AB}}|)$ are
$\lambda_{1}'=\frac{1}{2}(1-q)$ and
$\lambda_{2}'=\frac{1}{2}(1+q)$, we obtain
\begin{equation}\label{iaeeres}
 \mathcal{I}_{\mathrm{A:E}} =
\sum_{i=1}^{2} \lambda_{i}' \mathrm{log}\lambda_{i}'
 -\sum_{j=1}^{4} \lambda_{j} \mathrm{log}\lambda_{j}.
\end{equation}
As previously, Alice and Bob has only access to $\mathrm{P}_{01}$
and $\mathrm{P}_{10}$, thus we evaluate $\max_{c,d}
\{\mathcal{I}_{\mathrm{A:E}}\}$, the maximum value of
$\mathcal{I}_{\mathrm{A:E}}$ in the range of $c$ and $d$ allowed
values.

In Fig. 1 (b) $\max_{c,d} \{\mathcal{I}_{\mathrm{A:E}}\}$ is
plotted versus $\mathrm{P}_{01}$ and $\mathrm{P}_{10}$ where only
two regions are identified.
\begin{itemize}
  \item For $\mathrm{P}_{01}+ \mathrm{P}_{10} \geq 0.5$
   \\
$\max_{c,d}
\{\mathcal{I}_{\mathrm{A:E}}\}=\mathcal{I}_{\mathrm{A:E}}(\mathrm{P}_{01},
\mathrm{P}_{10}, c=1, d=-1 )$
 \item for $\mathrm{P}_{01}+ \mathrm{P}_{10} < 0.5$   \\
$\max_{c,d} \{\mathcal{I}_{\mathrm{A:E}}\}=1.$
\end{itemize}
Note that the maximum information achievable by Eve is
upper-bounded only if $\mathrm{P}_{01}+ \mathrm{P}_{10} \geq 0.5$,
but also in this case Eve cannot gain any information for
$\mathrm{P}_{01}=\mathrm{P}_{10}=0.5$. It is straightforward to
demonstrate that $\max_{c,d}
\{\mathcal{I}_{\mathrm{A(B):E}}\}(\mathrm{P}_{01}=\mathcal{P},\mathrm{P}_{10}=\mathcal{P})\geq
\max_{c,d}
\{\mathcal{I}_{\mathrm{A(B):E}}\}(\mathrm{P}_{01}=\mathcal{P}-k,\mathrm{P}_{10}=\mathcal{P}+k)$
with $-\mathcal{P} \leq k \leq \mathcal{P}$. Despite the different
shapes of the two surfaces in Fig. 1, for any values of
$\mathcal{P}$, we observe that $\max_{c,d}
\{\mathcal{I}_{\mathrm{A:E}}\}(\mathrm{P}_{01}=\mathcal{P},\mathrm{P}_{10}=\mathcal{P})=
\max_{c,d}
\{\mathcal{I}_{\mathrm{B:E}}\}(\mathrm{P}_{01}=\mathcal{P},\mathrm{P}_{10}=\mathcal{P})=H(1
- 2 \mathcal{P})$ for $0.25<\mathcal{P}\leq 0.5$, while
$\max_{c,d}
\{\mathcal{I}_{\mathrm{A:E}}\}(\mathrm{P}_{01}=\mathcal{P},\mathrm{P}_{10}=\mathcal{P})=
\max_{c,d}
\{\mathcal{I}_{\mathrm{B:E}}\}(\mathrm{P}_{01}=\mathcal{P},\mathrm{P}_{10}=\mathcal{P})=1$
for $0\leq\mathcal{P}< 0.25$ where $H$ is the Shannon entropy of a
binary channel \cite{NC00}.

In summary, the maximum information achievable by Eve is
upper-bounded in some region, even when Eve can modify the number
of photons in the quantum channel. Moreover these upper bounds
have been demonstrated to be strictly related to the
Anticorrelation Check outcomes. Thus measurement of
$\mathrm{P}_{01}$, $\mathrm{P}_{10}$ would allow Alice and Bob to
determine the security level of the communication. In addition we
point out that, as in \cite{qdkd}, Eve's resources have been
heavily overestimated in deriving Eq.s (\ref{ibeeres}) and
(\ref{iaeeres}). In order to extract information on Alice and Bob
operations, we assumed that Eve should perform any POVM on the
final state of the whole space $\mathcal{H}_{A} \otimes
\mathcal{H}_{B} \otimes \mathcal{H}_{E}$ as clearly stated in Eq.s
(\ref{ibee}) and (\ref{iaee}). This is obviously not the case.
Even if Eve can perform any POVM on the final state of her
\textit{ancilla} system, about the Alice-Bob system, she can only
know the results disclosed during the public discussion. This
induces to think that information achievable by Eve should be in
some cases well below these limits.

As already pointed out in \cite{qdkd}, Alice's and Bob's are able
to recover secure cryptographic keys in spite of Eve's attack if
the condition $I_{\mathrm{A:B}}> I_{\mathrm{A:E}.}$ and
$I_{\mathrm{A:B}}> I_{\mathrm{B:E}}$ is satisfied
\cite{ekertperes}, where $I_{\mathrm{A:B}}$ is the mutual
information between Alice and Bob (Ref. \cite{wolf} demonstrates
that this condition is, in some cases, too restrictive).
$I_{\mathrm{A:B}}$ can be simply calculated considering the
capacity of a noisy channel of quantum bit error rate
$\mathcal{Q}$, as $I_{\mathrm{A:B}}=1-H(\mathcal{Q})$ \cite{NC00}.
To ensure the security of the two generated keys, we replace
$I_{\mathrm{A:E}}$ and $I_{\mathrm{B:E}}$ with $\max_{c,d}
\{\mathcal{I}_{\mathrm{A:E}}\}$ and $\max_{c,d}
\{\mathcal{I}_{\mathrm{B:E}}\}$  calculated for
$\mathrm{P}_{01}=\mathrm{P}_{10}=\mathcal{P}$. This means that
Alice and Bob can distill common secret keys when $0.25 <
\mathcal{P} \leq 0.5 $ and
\begin{equation}\label{dis2}
  H(\mathcal{Q})+H(1-2\mathcal{P})<1.
\end{equation}

Despite the fact that Eq. (\ref{dis2}) appears to be formally
analogous to Eq. (6) in \cite{qdkd}, we underline that in Eq.
(\ref{dis2}) the term $\mathcal{P}$
($\mathcal{P}\simeq\frac{\mathrm{P}_{01}+\mathrm{P}_{10}}{2}$)
should be carefully evaluated from the experimental data as the
ratio between the anticorrelated results and all possible results
(anticorrelated and "wrong") of the Anticorrelation Check. The
results of the Anticorrelation Check should be considered "wrong"
not only in the case of correlated results, but also if more than
two photon are detected in coincidence by the Alice and Bob
apparatuses, or if only one of Alice's detector fires. This last
case makes the QDKD protocol not practical for today technology as
all the transmission losses and detection inefficiencies would be
considered due to eavesdropping attack.

Let us consider, as an example, the experimental realization of
QDKD protocol performed in \cite{qdkd}. Losses due to the
detection apparatuses (which is the main contribution) as well as
to the source and to the encoder apparatuses strongly affect the
estimation of the parameter $\mathcal{P}$. In fact, the
probability of losing one photon of the pair by the
generation-detection apparatuses was estimated to be $P_{loss}
\simeq 0.77$, and we observed a probability of correlated results
$P_{corr}<0.05$ due to optics imperfections, misalignment and dark
counts. The probability of measuring a three-fold or four-fold
coincidences due to the presence of more than a photon pair in the
quantum channels or to dark counts was completely negligible. Also
losses in the quantum channels are completely negligible
(propagation in air for less than one meter). Thus the
anticorrelation parameter $\mathcal{P}$ is evaluated as
$\mathcal{P}=(1-P_{loss}-P_{corr})/2 \simeq 0.09$, and, according
to the considerations related to Eq. (\ref{dis2}), common secret
keys cannot be distilled.

But if we assume that Eve cannot modify Alice and Bob detection
apparatuses as she has not access to their laboratories, the
detection inefficiencies can be traced out in the evaluation of
$\mathcal{P}$, and we could implement secure QDKD protocols for
short distances and in low noise environment. In particular, in
the case of the experimental implementation of Ref. \cite{qdkd}
the anticorrelation parameter simply reduces to
$\mathcal{P}=(1-P_{corr})/2 \simeq 0.47$, and the secure keys
distribution can be ascertained.

In Ref.s \cite{bostrom} and \cite{caili} two protocols are
proposed for unidirectional secure direct communication (message
from Bob to Alice) based on local operations on one photon of an
EPR pair. We observe that the QDKD scheme can be used to implement
bidirectional secure deterministic communication (either message
from Alice to Bob or from Bob to Alice).

When a deterministic message is sent from Bob to Alice, Alice
encodes a random sequence of bits and Bob encodes the
deterministic message. As Alice is aware of her operations
($\mathbf{1}_{\mathrm{B}}$ or $\widehat{Z}_{\mathrm{B}}$), she can
extract the message-bit encoded by Bob from the measurement result
$|\psi^{+}_{\mathrm{AB}}\rangle$ or
$|\psi^{-}_{\mathrm{AB}}\rangle$. During the communication Bob
discloses the results obtained from his Anticorrelation Check
apparatus on a non-jammable public channel, i.e. a public channel
that can be monitored but not modified by anybody else. According
to these results as well as to her measurements, Alice estimates
the security level of the on-going communication. If the
communication is insecure, Alice and Bob decide to abort the
transmission. A drawback of this protocol, as well as of the
protocols in Ref.s \cite{bostrom} and \cite{caili}, is that part
of the secret message is in any case eavesdropped before the
transmission is stopped.

When a deterministic message is sent from Alice to Bob, Bob
encodes a random sequence of bits (in some sense a cryptographic
key), and Alice encodes the deterministic message. The measurement
of a sequence of $|\psi^{+}_{\mathrm{AB}}\rangle$,
$|\psi^{-}_{\mathrm{AB}}\rangle$ corresponds to the Alice's
message encrypted by Bob's key. After the end of the quantum
communication, Bob discloses the results obtained from his
Anticorrelation Check apparatus on a non-jammable public channel.
According to these results as well as to her measurements, Alice
estimates the security level of the communication. If
communication is secure, Alice discloses the results of her
measurements corresponding to the message encrypted by Bob's key.
Since Bob is aware of his key, he can extract the message encoded
by Alice. Only if the security of the communication is ascertained
the encrypted message is publicly disclosed by Alice, thus this
protocol is not affected by the security drawback present in the
communication from Bob to Alice and in protocols of Ref.s
\cite{bostrom} and \cite{caili}.

Furthermore in both protocols of Ref.s \cite{bostrom} and
\cite{caili} the security proofs consider that Eve's presence only
induces bit-flips and they do not consider eavesdropping
strategies based on injection-subtraction of photons in the
quantum channel. In fact a successful eavesdropping attack against
protocol in Ref. \cite{bostrom} based on subtraction of photons
has already been proposed \cite{wojcik}, while valid eavesdropping
strategies have not yet been found against protocol of Ref.
\cite{caili}. We observe that for both these protocols a security
proof analogous to the one here proposed is necessary in order to
guarantee the security against also the attacks based on
injection-subtraction of photons.

In conclusion, this Reply proposes an improved security proof of
the QDKD protocol which is able to detect any individual
eavesdropping attack and to provide an upper bound to the
information achievable by Eve, even in the case of attacks
exploiting the possibility of injecting-subtracting photons in the
quantum channel as the ones proposed in W\'{o}jcik's Comment
\cite{wojcikcomment}.

The work was supported by MIUR (Project 67679) and by Elsag S.p.A.

\newpage

%\thebibliography


\begin{thebibliography}{99}
\bibitem{note} We remind that the security proof presented in
\cite{qdkd} assumes that Eve's presence only induces bit-flip,
i.e.
$|\Gamma_{\mathrm{BE}}\rangle=|0^{1}_{\mathrm{B}}\rangle\otimes|\gamma_{\mathrm{E}}\rangle
$ and
$|\Delta_{\mathrm{BE}}\rangle=|1^{1}_{\mathrm{B}}\rangle\otimes|\delta_{\mathrm{E}}\rangle
$. This implies that in the case of \cite{qdkd} $c$ and $d$ have
fixed values ($c=1$ and $d=-1$), while in the case discussed in
this paper are free parameters.
\end{thebibliography}
\end{document}